\documentclass{pnastwo}

\usepackage[pdftex]{color,graphicx}
\usepackage[numbers,comma,sort&compress]{natbib}
\usepackage{amssymb,amsfonts,amsmath}

\advance\voffset -8mm      

\newif\ifworkingversion

\workingversiontrue

\ifworkingversion
	\newcommand{\wordcount}[1]{}
\fi

\newcommand{\ket}[1]{\mbox{$\vert #1 \rangle$}}

\newsavebox{\upstate}
\newsavebox{\upstatesmaller}
\newsavebox{\downstate}
\newsavebox{\downstatesmaller}

\savebox{\upstate}{$\ket{\!\uparrow}$}
\savebox{\downstate}{$\ket{\!\downarrow}$}
\savebox{\upstatesmaller}{\raisebox{-2pt}{\scalebox{0.7}{$\ket{\!\uparrow}$}}}
\savebox{\downstatesmaller}{\raisebox{-2pt}{\scalebox{0.7}{$\ket{\!\downarrow}$}}}

\url{www.pnas.org/cgi/doi/10.1073/pnas.1204285109}
\copyrightyear{2012}
\issuedate{June 19, 2012}
\volume{109}
\issuenumber{25}
\footlineauthor{Steffen et al.}
\begin{document}

\title{Digital atom interferometer with single particle control on a discretized spacetime geometry}

\author{%
	Andreas Steffen\affil{1}{Institut f\"ur Angewandte Physik, Universit\"at Bonn,
	Wegelerstr.~8, D-53115 Bonn, Germany},
	Andrea Alberti\affil{1}{},
	Wolfgang Alt\affil{1}{},
	Noomen Belmechri\affil{1}{},
	Sebastian Hild\affil{1}{},
	Micha\l\ Karski\affil{1}{},
	\mbox{Artur Widera\affil{2}{Fachbereich Physik und Forschungszentrum OPTIMAS, Universit\"at Kaiserslautern, Erwin-Schr\"odinger-Stra\ss{}e,
	D-67663 Kaiserslautern, Germany}},%
\and
	Dieter Meschede\affil{1}{}}  

\contributor{Submitted to Proceedings of the National Academy of Sciences
of the United States of America}

\maketitle

\begin{article}
\begin{abstract}
	Engineering quantum particle systems, such as quantum simulators and quantum cellular automata, relies on full coherent control of quantum paths at the single particle level. Here we present an atom interferometer operating with single trapped atoms, where single particle wave packets are controlled through spin-dependent potentials. The interferometer is constructed from a sequence of discrete operations based on a set of elementary building blocks, which permit composing arbitrary interferometer geometries in a digital manner. We use this modularity to devise a spacetime analogue of the well-known spin echo technique, yielding insight into decoherence mechanisms. We also demonstrate mesoscopic delocalization of single atoms with a ``separation-to-localization'' ratio exceeding 500; this suggests their utilization beyond quantum logic applications as nano-resolution quantum probes in precision measurements, being able to measure potential gradients with precision $\mathbf{5\times 10^{-4}}$ in units of gravitational acceleration~$\boldsymbol{g}$.
	\wordcount{158}
\end{abstract}


% \abbreviations{SAM, self-assembled monolayer; OTS, octadecyltrichlorosilane}

\begin{figure*}[t]
\ifworkingversion
\centerline{\includegraphics[width=\textwidth]{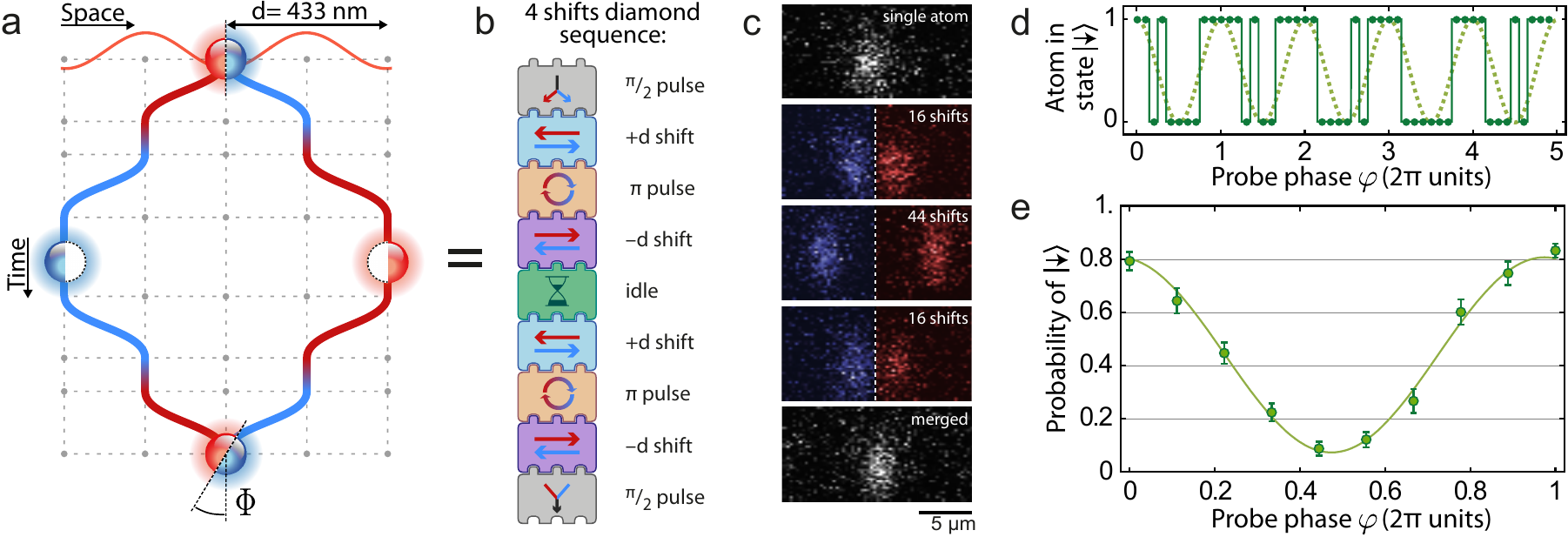}}
\fi
\caption{{Digital single atom interferometer.} \textbf{a)} Basic ``diamond'' geometry: an atom is split in a spin state superposition, the two spin components are then coherently shifted according to their state, and eventually merged to extract the phase difference $\Phi$ accumulated between them.
Shift operations of alternating displacement $\pm{}d$ are interleaved with $\pi$ pulses.
\textbf{b)} Digital representation in elementary operation blocks.
Assembling block sequences permits modular interferometer geometries; idle times can extend the interrogation time.
\textbf{c)} Single atom images illustrating coherent delocalization at different separations.
The dashed lines separate two independently recorded images; their initial atom positions are overlapped with single-site precision \cite{karski2009nearest}.
The localization of 18 nm RMS is not visible due to diffraction-limited optical resolution.
\textbf{d)} With the sequence in b), detecting $\ket{\!\downarrow}$ state by interrogating one atom at a time for different probe phase $\varphi$ (defined in text) results in a binary signal.
The dotted sine idealizes the underlying probability distribution to guide the eye.
\textbf{e)} Averaging over almost 2000 atoms results in a Ramsey interference fringe reflecting the probability to detect $\ket{\!\downarrow}$.
The interferometric phase $\Phi$ is extracted by fitting the fringe to Eq.~\ref{eqSine}. Error bars show one SEM uncertainty.
\wordcount{189}}
\label{fig1}
\end{figure*}

\section{Localized particles}
\dropcap{P}article interferometry has yielded numerous important results from the advent of quantum mechanics until now; counting only the most recent ones, they range from precision measurements \cite{peters1999measurement}, fundamental tests of quantum mechanics \cite{arndt1999wave,Nesvizhevsky:2002dl}, to applications in optical and superconducting magnetometry \cite{2007NatPhBudker,BraginskiBook}. While other architectures provide a natural way to engineer, guide and confine interfering paths, e.g., by means of optical waveguides for photons or superconducting circuits for electrons, one key challenge for atoms consists in realizing a flexible platform that is capable of providing complex interfering geometries where the atoms remain localized at all times. Unprecedented control over atomic states has been obtained in optical lattice potentials: designed to mimic solid state systems they resemble engineered circuits. In order to construct interfering paths in these systems, an additional quantum number is needed -- that has to be decoupled from motional states -- for the atomic wave packet's motion to be steered in a coherent and state-dependent manner. This is achieved by leaving the simple scalar concept of an atom and taking advantage of its vector nature endowed by electronic and nuclear spin. Mediated by the spin-orbit interaction, optical dipole forces yield a state dependent force, offering the right ``handle'' for steering atomic paths. Therefore, an effective magnetic field coupled to the atomic magnetic moment makes position control of the interfering paths possible, albeit at the expense of an increased sensitivity to magnetic fields \cite{Deutsch:1998jn,Brennen1999Quantum,jaksch1999entanglement}. Even though this is not a fundamental limitation, it represents a technical hurdle that demands efficient shielding against stray magnetic field fluctuations; accepting this price, digital engineering of atomic paths in optical potentials can bring atom interferometry to a new level of control.

Particle localization and digital reprogrammability of atomic paths, besides being necessary resources in quantum technologies, e.g., for quantum cellular automata \cite{Grossing:1988p3510} and multi-particle entanglement \cite{mandel2003controlled}, specify also the application range where trapped-atom interferometers could outperform other interferometers in precision measurements.
In general, to measure constant homogeneous potential gradients the figure of merit is the space-time area enclosed by the atomic paths, and the ultimate precision is fixed by shot noise as standard quantum limit. While space-time areas achieved by free-fall atom interferometers, e.g., of about $1\,\textrm{mm}\times 160\,\textrm{ms}$ in \cite{peters1999measurement}, are in principle reachable by our interferometer, and at the moment only technically limited, the prime advantage in having trapped atoms resides in the nanoscale spatial resolution, allowing non-destructive probing of microscopic atomic quantum systems and measuring potential gradients at ultrashort scale.
\wordcount{436}

\section{The interferometer}
Atom interferometers operate by splitting the wavefunction of an atom onto two spatially separated paths, and measuring their phase difference \cite{berman1997,cronin2009optics}.
As depicted in Fig.\ \ref{fig1}a this requires preparing the atom in a quantum superposition of states; these are subsequently delocalized and eventually coherently recombined to probe the phase difference accumulated between the two paths.
Our interferometer uses two long-lived internal states of cold Caesium atoms,  $\ket{\!\uparrow}=\ket{F=4,m_F=4}$ and $\ket{\!\downarrow}=\ket{F=3,m_F=3}$. While these internal states are coherently manipulated by means of microwave pulses, the particle position is controlled by trapping atoms in a 1D\ optical lattice with spacing $d=433$~nm.
The spin-dependent control is realized through a $\textrm{lin}$-$\theta$-$\textrm{lin}$ configuration of the optical lattice, which allows atoms to move in discrete steps in a direction dependent on the spin state
$\ket{\!\uparrow,x} \longrightarrow \ket{\!\uparrow,x\pm d/2} $, $\ket{\!\downarrow,x} \longrightarrow \ket{\!\downarrow,x\mp d/2}$ by ramping the polarization angle $\theta$ (see Methods) \cite{mandel2003coherent,Lee:2007ip}.
When a superposition of the two spin states is created by a $\pi/2$ pulse, the trapped atom is delocalized by a single step in a fully coherent manner. As illustrated in Fig.\ \ref{fig1}b, shift operations of alternated direction are interleaved with $\pi$ pulses to further separate the paths. Up to 100 block operations can be concatenated in one coherent digital chain, eventually achieving coherent separations larger than 10 $\mu$m, as shown in Fig.\ \ref{fig1}c.
The tight confinement by the lattice potential ensures that each wavefunction's component is axially strongly localized in space, down to $18$ nm, achieving at the same time nanoscale positioning control and mesoscopic spatial separation.

In order to extract the interferometric phase difference $\Phi$ between the two paths, we employ a Ramsey probing scheme where the final merging block maps the phase information onto the two spin populations by means of a $\pi/2$ pulse of variable phase $\varphi$. Scanning the phase varies the probability to detect $\ket{\!\downarrow}$
\begin{equation} 	
\label{eqSine} p_{\usebox{\downstatesmaller}}(\varphi)=\frac{(1-\gamma)}{2} \cdot \left[1 + C\cdot \cos\left(\Phi + \varphi\right)\right],
\end{equation}
$\gamma$ being the probability of an atom escaping from the trap during the interferometer sequence, and $C$ the Ramsey fringe contrast.
When the interferometer is operated with a single atom, detecting the presence of the atom in, e.g., $\ket{\!\downarrow}$ state results in a $0/1$ binary signal at each run of the sequence, as displayed in Fig.\ \ref{fig1}d.
Unlike with large ensembles of atoms, to measure the underlying probability distribution in Fig.\ \ref{fig1}e  multiple repetitions are here averaged in time.
\wordcount{368}

\section{Geometries}
The geometry of the interferometer is determined by the programmed sequence of operation blocks.
The elementary ``diamond'' geometry, which resembles the classic Mach-Zehnder interferometer, maximizes the sensitivity of the interferometric phase to potential gradients.
Varying the size of the diamond reveals how phase and coherence of the superposition state evolve with the number of shifts.
Fig.\ \ref{fig2}a shows a strong parabolic phase accumulation, which we can track over 11\,$\pi$ radians.
Such a phase behavior is expected in the presence of a linear potential gradient  $\nabla U$, as it can be interpreted as the gradient strength times the spacetime ``area'' that the paths enclose.
The accumulated phase will then follow
\begin{multline}
	\label{eqDiamPhase}
\Phi(n) = \frac{1}{\hbar}\int U(x_L(t))\, \textrm{dt} -\frac{1}{\hbar}\int U(x_R(t))\, \textrm{dt} =\\ =\frac{\nabla{}U}{\hbar} \cdot d\,\left[\left(\frac{n}{2}\right)^{\hspace{-1.5pt}2}\hspace{-1.5pt}\cdot (\tau_S + \tau_\pi)- \frac{n}{2} \cdot \tau_\pi\right],
\end{multline}
$n$ being the total number of shift operations, $\tau_S$ the duration of a shift block, and $\tau_\pi$ the duration of a $\pi$ pulse, and $x_L$, $x_R$ the coordinate along the left and right path, respectively.
The phase evolution in Eq.~\ref{eqDiamPhase} has been fitted to Fig.\ \ref{fig2}a, indicating a potential gradient of $2\pi\hbar\times (324.5 \pm 0.8)\,\mathrm{Hz}/d$, equivalent in magnitude to a gravitational acceleration of $(0.2296 \pm 5\cdot10^{-4})\,g$.
We attribute this gradient to a displacement of the lattice laser focus from the position of the atoms by $\approx 600\, \mu$m, about a quarter of the Rayleigh length;
the beam's divergence leads to a reduction of the optical trap depth further away from the focus, thus creating a potential gradient which is nearly linear over the experimental region of $\approx 40\,\mu$m.
This is supported by the direct proportionality measured between the gradient strength and the trapping laser's power, as shown in the inset of Fig.\ \ref{fig2}a.

In order to cancel the phase caused by such a background gradient, we consider a ``double diamond'' geometry, consisting of two basic diamond interferometers interspaced with a $\pi$ pulse, see  Fig.\ \ref{fig2}b.
By crossing the paths in the middle of the sequence, the second loop accumulates a phase opposite to the first loop, and cancellation occurs akin to a spin echo.
This ``geometric'' spin echo is effective for any spin- and time-independent potential.
Gradients which are active only for one half of the sequence are still detected as described above, making the double diamond interferometer suitable to measure time-controllable gradients in a differential scheme.
The successful suppression of the parabolic phase is visible in Fig.\ \ref{fig2}a.
\wordcount{392}

\section{Decoherence}
\begin{figure*}[b]
\ifworkingversion
\centerline{\includegraphics[width=\textwidth]{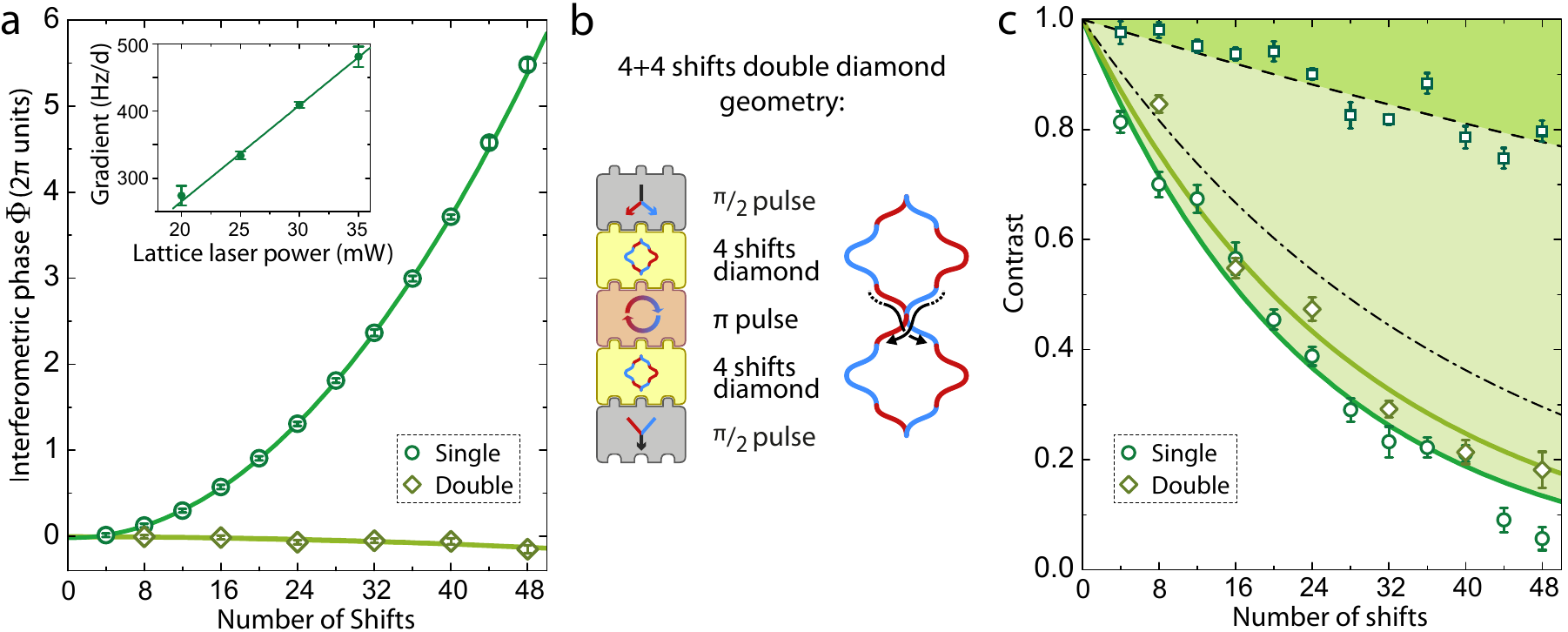}}
\fi
\caption{{Interferometric phase and contrast vs. paths' length.}
\textbf{a)} The phase accumulation reveals potential gradients (single diamond geometry).
The parabolic behavior indicates a linear potential gradient due to spacetime area's quadratic dependence on the number of shifts operations; it is verified to be proportional to the lattice laser power (inset).
The fit to Eq.~\ref{eqDiamPhase} determines the local force with precision $5\times 10^{-4}$ in units of $g$, corresponding to $10^4$ atoms sequentially interrogated in $90$ minutes.
A geometric analogue of spin echoes suppresses the accumulated phase (double diamond geometry). The recorded gradient is reduced by a factor of 10, where interferometers with equal maximal separation are compared, e.g., 12 shifts single diamond with 24 shifts double diamond.
\textbf{b)} The double diamond is composed of two single diamonds interspaced with a $\pi$ pulse;
path crossing in the middle yields a geometric cancellation of the phase.
\textbf{c)} The fringe contrast decays exponentially with the number of shifts.
The dashed line is the exponential fit to the fringe contrast measured with shift operations replaced by idle blocks; it reflects the spin coherence time.
Correcting it for the near-unity shift fidelity yields the dash-dotted line.
The increased contrast in the double diamond is likely due to geometric echo refocusing.
Error bars show one SEM uncertainty.
\wordcount{198}}
\label{fig2} 
\end{figure*}

To mitigate decoherence effects the interferometer block chain is assembled such that each subsequence of shift, $\pi$ pulse, shift operations achieves on its own a spin echo refocusing, compensating spin-dependent inhomogeneous disturbances; such repeated spin echoes are reminiscent of dynamical decoupling \cite{Viola:1998p3485,Andersen:2004p3541}.
Fig.\ \ref{fig2}c shows an exponential decay of contrast for increasing interferometer size, which we observed for both interferometer geometries. This behavior indicates that identically repeated operation blocks are subject to an invariable decoherence per block, suggesting the number of shifts as the relevant quantity for contrast. 

To isolate the cause of this decay, we repeat the same sequence replacing all the shift operations by idle blocks, and fit the contrast decay with an exponential curve (dashed line) showing a contrast loss of 0.6\% per step. This is dominated by time-varying differential light shift due to fluctuations of the lattice laser power; in addition, since spin-dependent transport excludes the use of clock states, a minor contribution from magnetic field fluctuations is also expected.

The significantly lower decay rate without delocalization on the lattice (dark green area) indicates that most decoherence in the interferometer takes place due to shift operations (light green area). We hence measure independently the transport efficiency per single shift step $\pm{}d/2$ using methods of \cite{2011karski}, and, after preparing atoms in the axial motional ground state \cite{foerster2009microwave}, we achieve 99\% success probability for shifting the atom in the correct direction, limited by the efficiency of the microwave pulses.
Accounting for this probability once per shift and per path, the shift-free decay rate is combined with a further 2\% contrast drop per step to produce the dash-dotted curve in Fig. 2c. We attribute the residual 1.7\% contrast drop per step to polarization jitter during shift operations and, to a smaller extent, to spurious motional excitations.
\wordcount{292}

\section{Hold times}
\begin{figure}[t]        
\ifworkingversion
\centerline{\includegraphics[width=1\columnwidth]{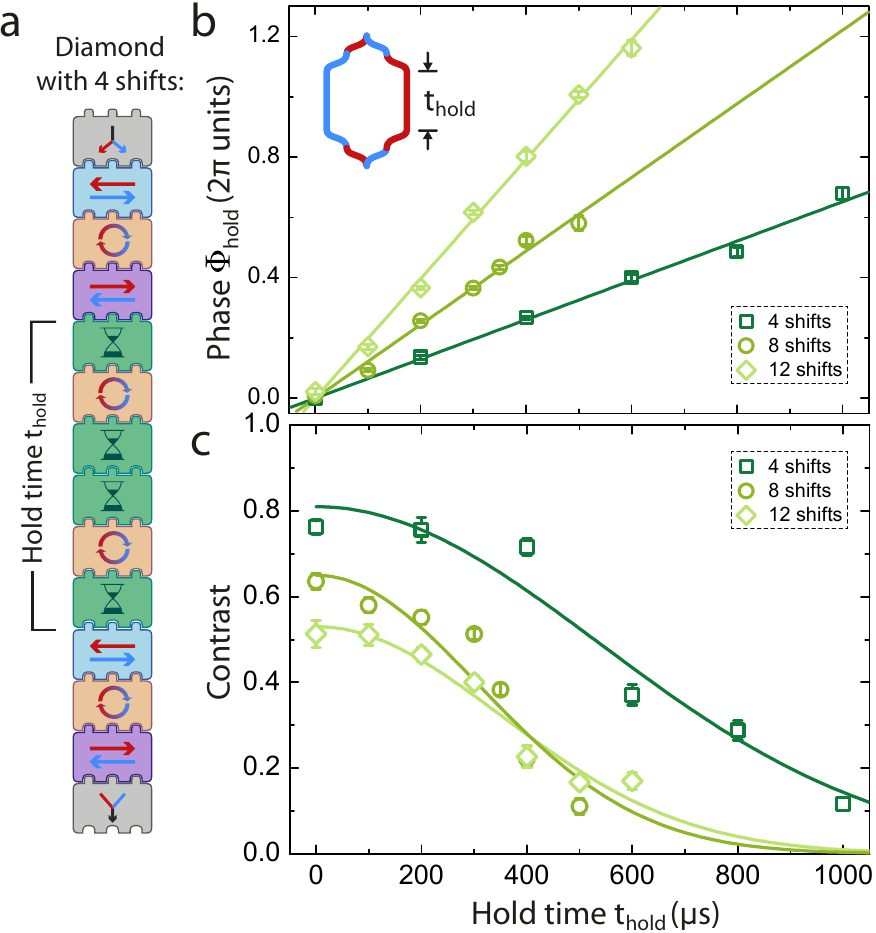}}
\fi
\caption{{Hold time in a diamond geometry.}
\textbf{a)} Digital representation: two $\pi$ pulses during the hold time form a double spin-echo sequence;
all idle blocks have identical duration.
\textbf{b)} The hold time contributes an additional phase $\Phi_\mathrm{hold}$, increasing linearly with $t_\mathrm{hold}$ with the slopes proportional to path separation.
The intercepts of the fitting lines have been subtracted to set the phase to zero in the origin. \textbf{c)} The contrast reveals a Gaussian decay with hold time. This dependency on the echo duration is characteristic of homogeneous dephasing mechanisms which are not compensated by spin echo pulses \cite{kuhr2005analysis}; it differs from the exponential decay in Fig.\ \ref{fig2}c where the number of echo pulses is varied. Error bars show one SEM uncertainty.
\wordcount{90}}
\label{fig3} 
\end{figure}

One distinguished feature of a trapped particle interferometer is the ability to insert an arbitrary hold time $t_\mathrm{hold}$ in between blocks, where the interferometer is held open at a fixed separation. This becomes relevant when the interferometer is to be operated near an object of study, with long interaction time and nanoscale control of distance, e.g., a surface or even a Bose-Einstein condensate.
The natural place to insert a hold time is at maximum separation of the paths, see Fig.\ \ref{fig3}a; this situation is investigated in Fig.\ \ref{fig3}b for sequences of $n=4$, 8 and 12 shifts where the interferometric phase is measured varying $t_\mathrm{hold}$. In the presence of a potential gradient an additional phase $ \Phi_\mathrm{hold}=\nabla U \cdot n\,d\,t_\mathrm{hold}/(2\hbar)$ arises from the spacetime area added by the hold time, on top of the phase in Eq.~\ref{eqDiamPhase}.
A linear fit yields a potential gradient of $2\pi\hbar\times (324 \pm 7)$ Hz/$d$, in full agreement with the results in Fig.\ \ref{fig2}.

To protect the superposition state from dephasing during the hold time two $\pi$ pulses are applied, realizing a two-fold spin echo refocusing, see Fig.\ \ref{fig3}a.
The fact that the phase $\Phi_\mathrm{hold}$ survives spin echoes confirms the spin-independent nature of the potential gradient, as expected from such a light shift gradient.
As shown in Fig.\ \ref{fig3}c, the contrast as a function of the hold time reveals a Gaussian decay \cite{kuhr2005analysis} on a time scale comparable to that of non-transported atoms.
\wordcount{287}

\section{Inertial force measurement} 
An important application of atom interferometry is the measurement of external forces. 
We demonstrate this by applying a controlled inertial force by accelerating the whole lattice potential during a fixed acceleration time $t_\mathrm{acc}$ (see Methods).
As shown in Fig.\ \ref{fig4}a, the paths are then recombined in the moving inertial frame, and the effect of the force $m\hspace{1pt}a$ is measured from the interferometric phase, with $m$ the atomic mass and $a$ the applied acceleration.
We employ the simplest geometry of the single diamond and find that the phase linearly increases with the applied acceleration $a$ with a slope proportional to path separation, see Fig.\ \ref{fig4}b.
Calculating the spacetime area in this case yields a phase of $\Phi_\mathrm{acc} = m\hspace{1pt}a\cdot\hspace{1pt}n\hspace{1pt}d \hspace{1pt}t_\mathrm{acc}/(2\hbar)$ for $n$ total shifts, where $a$ is independently calibrated. The agreement up to $5\,g$ -- only technically limited -- together with the achieved precision of $5\times 10^{-4}g$ in Fig.\ \ref{fig2} demonstrate the ability of the single atom interferometer to measure external fields in real applications, showing a dynamic range of at least 4 orders of magnitude.
% Owing to its single-atom nature this interferometer does not aim at competing with $10^{-9}$ relative precision demonstrated, e.g., by Peters \emph{et al.}\ \cite{peters1999measurement}, but rather at exploiting the tight localization to investigate potential gradients at ultrashort scales.
\wordcount{166}

\section{Conclusions and outlook}
In summary, we have demonstrated an atom interferometer operating at the single atom level; we proved its capability to measure external fields -- both the divergence of an optical dipole trap and an externally-controlled inertial force -- achieving a resolution which borders precision applications. By exploiting the block modularity we could precisely identify the decoherence mechanisms and explain the loss of contrast in terms of decoherence per shift operation. Yet, the full digital control established here allowed us to suppress decoherence effects by engineering the interferometer's paths to effect a geometrical refocussing.

In the quest for quantum devices functioning at the single-atom level, from Rydberg quantum logic gates \cite{Urban:2009jd,Gaetan:2009fq} to single-site-resolved optical lattices \cite{karski2009nearest,Bakr:2009p606,weitenberg2011single}, our results lay the base for digital quantum simulators \cite{Lloyd:1996p3374} and quantum cellular automata \cite{Grossing:1988p3510,lloyd1993} based on interacting atoms in lattice potentials \cite{karski2009quantum,Ahlbrecht:2011p3404}. Similarly, scalable arrays of quantum logic gates with single atom addressing are also within reach \cite{mandel2003controlled}.
Further, this system promises to reveal quantum coalescence of massive boson particles in a two-atom Hong-Ou-Mandel interference \cite{Hong:1987p2009}, with noticeable implications for sub-shot-noise interferometry \cite{Giovannetti:2004p3506,Dunningham:2005p3465}.

The nanoscale single-atom control evidenced here is particularly suitable for the investigation of surface effects, such as Casimir-Polder interactions or Yukawa-like gravitational potentials at micrometer distances \cite{Adelberger:2003p1080}. Reducing technical sources of dephasing, we expect 10-fold improvement of coherence time, allowing measurement of Casimir-Polder potential with single atoms with 5\% precision at distances shorter than $4\,\mathrm{\mu m}$ from a dielectric surface, assuming 24 hours integration time \cite{Beaufils2011}. Furthermore, an implementation of conveyor belts for spin-dependent optical lattices is currently underway, with which we expect to reach coherent splitting distances of the order of $1\,\mathrm{mm}$ \cite{Kuhr:2001}. Combining this with a hold-time scheme for interrogation in decoherence-protected clock states, we deem realistic an improvement of space-time area by a factor $10^5$ over the present situation, eventually rivaling in precision free-fall atom interferometers.
\wordcount{290}

\begin{figure} 	
\ifworkingversion
\centerline{\includegraphics [width=1\columnwidth]{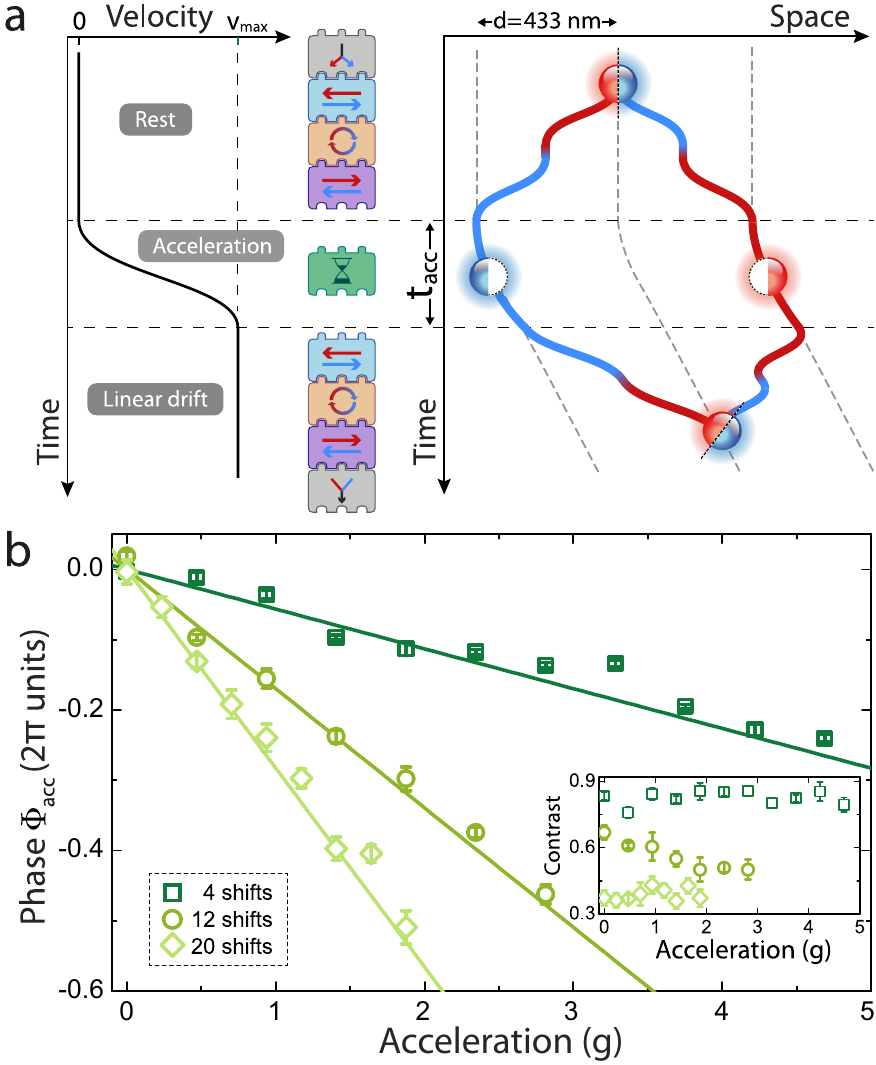}}
\fi
\caption{{Interferometric measurement of inertial forces.}
\textbf{a)} An inertial force is applied during the hold time by accelerating the lattice by up to $5\,g$ for a duration of $t_\mathrm{acc}=20\,\mu$s;
the paths are merged while in the inertial frame at velocity $v_\mathrm{max}$.
\textbf{b)} Phase $\Phi_\mathrm{acc}$ shows a linear dependence on acceleration for 4, 12 and 20 shifts.
The solid lines show the prediction by a simple model (see text) without free parameters where the acceleration exerted by means of a piezoelectric transducer has been independently calibrated.
Systematic deviations from the linear model are expected to occur because of mechanical resonances and nonlinearities in the response of the transducer.
Phase offsets of the data sets have been subtracted. Inset: constant contrast at given separation indicates that acceleration causes no decoherence. Error bars show one SEM uncertainty.
\wordcount{98}} 	
\label{fig4} 
\end{figure}

\begin{materials}
	\section{Lattice}
	Two counterpropagating linearly-polarized laser beams at wavelength $\lambda=866\,\mathrm{nm}$, with the polarization of the retroreflected beam rotated by a variable angle $\theta$ (lin-$\theta$-lin configuration), produce an optical lattice potential with spacing $d=\lambda/2$; the beams are focused to the position of the atoms with a Rayleigh length of $2.3\,\mathrm{mm}$.
	The trap frequencies of each potential well are $\omega_{ax}=2\pi \times 120\,\mathrm{kHz},\, \omega_{rad}=2\pi \times 2\,\mathrm{kHz}$, and the off-resonant scattering rate is $10\,\mathrm{Hz}$.
	After molasses cooling to $T=10\,\mathrm{\mu{}K}$, atoms are cooled in the lattice to the motional ground state along the axis \cite{foerster2009microwave}.
	The two spin states, $\ket{\!\uparrow}$ and $\ket{\!\downarrow}$, experience different light shifts induced by the left- and right-circular components of the laser; this allows spin-dependent shift operations by controlling the angle $\theta$ with an electro-optic modulator \cite{Brennen1999Quantum,jaksch1999entanglement,mandel2003coherent}. The beams' polarization is controlled with a relative precision of $10^{-4}$, and the shift duration is set depending on $\omega_{ax}$ to suppress motional excitations \cite{mandel2003coherent}.

	\section{Detection}
	The interferometric phase is mapped to spin populations by a $\pi/2$ pulse around the variable axis $\hat{\sigma}_x \cos\varphi + \hat{\sigma}_y \sin\varphi$. The spin population $p_{\usebox{\downstatesmaller}}(\varphi)$ is detected by selectively pushing out the atom when in the state $\ket{\!\uparrow}$, and subsequently detecting the presence of an atom by fluorescence image; atom losses $\gamma\approx{}5\%$ due to background collisions and light scattering during fluorescence imaging cannot be distinguished from state $\ket{\!\uparrow}$.

	\section{Coherence time} The bare spin coherence time is $T_2 \approx 200\,\mu$s. This originates mostly from radial thermal motion causing inhomogeneous variations of the differential light shift of the two hyperfine states \cite{kuhr2005analysis}. One spin echo pulse is able to suppress such an inhomogeneous dephasing, resulting in an extended coherence time $T_2^* \approx 600\,\mu$s. Spin echo series as in the interferometer sequence benefit from a dynamical decoupling effect \cite{Viola:1998p3485,Andersen:2004p3541}, yielding a longer $T_2^\star$ up to $2.3\,$ms. This time should be compared to the single step duration of $\tau_S=18\,\mathrm{\mu s}$ per shift plus $\tau_\pi=12\,\mathrm{\mu s}$ per $\pi$ pulse.

	\section{Scalability} 
	To increase the acquisition rate several interferometers are executed in parallel by loading on average about 30 atoms into the dipole trap over a region of $40\,\mathrm{\mu m}$ (Fig.\ \ref{fig2}, \ref{fig3}, \ref{fig4}), where each run of the interferometer lasts about $1.5\,\textrm{s}$. Because of the weak radial confinement, atom-atom interactions are negligible.

	\section{Inertial force}
	An inertial force is created by accelerating the lattice potential; we apply a parabolic voltage ramp to a piezo actuator holding the retroreflecting mirror. The displacement-to-voltage dependence is calibrated with a Michelson interferometer, producing the mean acceleration $\int\!a(t)\,\mathrm{d}t/t_\mathrm{acc}$ used in phase calculations. Methods from \cite{briles2010simple} raise the first piezo resonance to over 50 kHz; maximal acceleration is bounded by electronic limitations. Interband Landau-Zener tunneling is negligible below the critical acceleration of $50\times 10^3\,\mathrm{m}/\mathrm{s}^2$.
	\\[1mm]\noindent\wordcount{403}
\end{materials}

\begin{acknowledgments}
	We acknowledge support from the Deutsche Forschungsgemeinschaft FOR635, the European Commission's SCALA and AQUTE projects. A.~Steffen acknowledges support from the Studien\-{}stiftung des deutschen Volkes. A. Alberti acknowledges support by the A.~v.~Humboldt Foundation. A.~Steffen and S.~Hild acknowledge support from Bonn-Cologne graduate school. We thank Reinhard Werner for discussions.
\end{acknowledgments}

\end{article}

\end{document}